\documentclass[aps,twocolumn]{revtex4}

\usepackage{amsmath}
\usepackage{graphicx}

\usepackage{epsfig}

\def\ov#1{\overline{#1}}

\def\pd#1#2{\frac{\partial #1}{\partial #2}}

\def\cal#1{\mathcal{#1}}

\newcommand{\bc}{\begin{center}}
\newcommand{\ec}{\end{center}}
\newcommand{\bt}{\begin{tabbing}}
\newcommand{\et}{\end{tabbing}} 
\newcommand{\be}{\begin{eqnarray*}}
\newcommand{\ee}{\end{eqnarray*}}
\newcommand{\bs}{\begin{slide}}
\newcommand{\es}{\end{slide}}

\begin{document}

\begin{flushright}
November 12, 2010
\end{flushright}

\title{Compact formulas for guiding-center orbits in axisymmetric tokamak geometry}

\author{Alain J.~Brizard}
\affiliation{Department of Chemistry and Physics, Saint Michael's College \\ Colchester, VT 05439, USA} 

\begin{abstract}
Compact formulas for trapped-particle and passing-particle guiding-center orbits in axisymmetric tokamak geometry are given in terms of the Jacobi elliptic functions and complete elliptic integrals. These formulas can find applications in bounce-center kinetic theory as well as neoclassical transport theory.
\end{abstract}


\maketitle

\section{Introduction}

The presence of trapped-particle and passing-particle guiding-center orbits in axisymmetric tokamak geometry has had a significant impact on magnetic fusion research (see Refs.~\cite{Galeev_Sagdeev,Hinton_Hazeltine} for a brief historical survey). Because of their fundamental importance in understanding neoclassical and anomalous transport in tokamak plasmas \cite{anomalous}, it is desirable to find compact analytic representations for the trapped-particle and passing-particle guiding-center orbits in axisymmetric tokamak geometry. In particular, compact expressions might allow explicit calculations of bounce-angle averages in bounce-center kinetic theory \cite{GD_90,Brizard_bc,Wang_Hahm} beyond the deeply-trapped approximation. In addition, a bounce-center Fokker-Planck collision operator might be constructed from the guiding-center Fokker-Planck collision operator 
\cite{Brizard_gcFP}, which would enable a kinetic description of neoclassical transport beyond the zero-banana-width limit \cite{Brizard_OAgcFP}.

Previous analytic representations have used a simplified tokamak geometry \cite{RBW} in which the magnetic surfaces are unshifted concentric circular toroidal surfaces centered on the magnetic axis (located at a major radius $R$ for the axis of symmetry). In this simplified tokamak geometry, trapped-particle and passing-particle guiding-center orbits \cite{Wang_Hahm} are represented in terms of complete and incomplete elliptic integrals 
\cite{HMF_elliptic}. The purpose of this work is to present compact analytic formulas for these trapped and passing guiding-center orbits expressed in terms of the Jacobi elliptic functions \cite{HMF_Jacobi,Brizard_elliptic}. These compact formulas are readily amenable to explicit calculations that do not require using simplifying limits such as the deeply-trapped limit (in which case the Jacobi elliptic functions become simple trigonometric functions).

The remainder of the paper is organized as follows. In Sec.~\ref{sec:simple}, we introduce the simple axisymmetric tokamak geometry used here and in previous works \cite{RBW}. In Sec.~\ref{sec:action}, we introduce the bounce and transit actions $J_{\ell} = (J_{\rm b}, J_{\rm t})$ associated with the trapped-particle $(\ell = {\rm b})$ and passing-particle $(\ell = {\rm t})$ guiding-center orbits and their associated orbital frequencies $\omega_{\ell} \equiv (\partial J_{\ell}/\partial{\cal E})^{-1}$, which are given in terms of complete elliptic integrals \cite{Wang_Hahm}. In Sec.~\ref{sec:angle}, we present compact expressions for the poloidal angle $\vartheta$ in terms of Jacobi elliptic functions depending on the bounce angle $\zeta_{\rm b}$ (for trapped-particle orbits) and the transit angle $\zeta_{\rm t}$ (for passing-particle orbits). In Sec.~\ref{sec:canonical}, we introduce the canonical parallel coordinates $s(\zeta_{\ell}, J_{\ell})$ and $p_{\|}(\zeta_{\ell}, J_{\ell})$ for trapped-particle and passing-particle guiding-center orbits, which satisfy a canonical condition $\{s,\; p_{\|}\} \equiv 1$. This canonical conditions forms the basis of the bounce-center phase-space transformation 
\cite{Duthoit_etal} in general magnetic geometry. The fact that the parallel coordinates explicitly satisfy the canonical conditions is a major result of the present work. Lastly, in Sec.~\ref{sec:toroidal}, the slow toroidal-drift guiding-center dynamics is described in terms of compact expressions involving the Jacobi elliptic functions and our work is summarized in Sec.~\ref{sec:sum}.

\section{\label{sec:simple}Simple Tokamak Geometry}

In circular, large-aspect-ratio tokamak geometry \cite{RBW}, the magnetic-field strength is approximated as
\begin{eqnarray} 
B(\psi,\vartheta) & \simeq & B_{0}\,\left(1 - \frac{r}{R}\,\cos\vartheta\right) \nonumber \\ 
 & = & B_{\rm e} \;+\; 2\,B_{0}\,\epsilon\;\sin^{2}\frac{\vartheta}{2},
\label{eq:B_tok}
\end{eqnarray}
where the small inverse aspect ratio 
\begin{equation}
\epsilon(\psi) \;\equiv\; r(\psi)/R \;\ll\; 1 
\label{eq:epsilon_0}
\end{equation}
is defined in terms of the minor radius $r(\psi)$ of a circular magnetic surface (labeled by $\psi$), and the poloidal angle $\vartheta$ is measured from the outside equatorial plane. In Eq.~\eqref{eq:B_tok}, $B_{0}$ denotes the magnetic-field strength on the magnetic axis (at $r = 0$) and $B_{\rm e} \equiv 
B_{0}\,(1 - \epsilon)$ denotes the magnetic-field strength on the (outside) equatorial plane. Lastly, we note that the magnitude of the toroidal magnetic is $B_{\rm tor} \simeq B$ while the magnitude of the poloidal magnetic field at the minor radius $r$ is $B_{\rm pol} \simeq \epsilon(\psi)\,
B_{0}/q(\psi)$, where $q(\psi)$ is the safety factor \cite{RBW}. In all expressions presented here, terms of order $\epsilon^{2}$ are systematically omitted (the symbol $\simeq$ is used whenever this approximation is first introduced).

On a single magnetic surface (at fixed $\epsilon$), the magnitude of the parallel momentum $|p_{\|}|(\vartheta) \equiv \sqrt{2m\,({\cal E} - \mu\,B)}$ of a guiding-center of mass $m$ (at fixed total energy ${\cal E}$, magnetic moment $\mu$, and magnetic flux $\psi$) is
\begin{eqnarray}
|p_{\|}|(\vartheta) & \simeq & \sqrt{2m\,({\cal E} - \mu\,B_{\rm e})}\;\sqrt{1 \;-\; \left( \frac{2\epsilon\,\mu B_{0}}{{\cal E} - \mu\,B_{\rm e}}\right) \sin^{2}\frac{\vartheta}{2}} \nonumber \\
 & \equiv & p_{\|{\rm e}}\;\sqrt{1 \;-\; \kappa^{-1}\;\sin^{2}\frac{\vartheta}{2}} \;\equiv\; mR_{\|}\;|\dot{\vartheta}|.
\label{eq:ppar_tok}
\end{eqnarray}
Here, the equatorial parallel momentum (on the low-field side at $\vartheta = 0$) of the guiding-center orbit is defined as
\begin{equation}
p_{\|{\rm e}} \;=\; \sqrt{2m\,({\cal E} - \mu\,B_{\rm e})} \;=\; 2\sqrt{\kappa}\;mR_{\|}\omega_{\|},
\label{eq:ppar_equat}
\end{equation}
where the bounce-transit parameter \cite{GD_90} is defined as
\begin{equation}
\kappa({\cal E},\mu,\psi) \;\equiv\; \frac{{\cal E} - \mu\,B_{\rm e}}{2\epsilon\;\mu B_{0}} \;=\; \frac{\xi_{\rm e}^{2}\,(1 - \epsilon)}{
2\epsilon\,(1 - \xi_{\rm e}^{2})},
\label{eq:kappa_def}
\end{equation}
and $\xi_{\rm e} \equiv p_{\|{\rm e}}/p$ denotes the equatorial pitch-angle coordinate. In Eq.~\eqref{eq:ppar_equat}, we also defined the characteristic parallel frequency 
\begin{equation}
\omega_{\|} \;\equiv\; \frac{1}{R_{\|}}\;\sqrt{\epsilon\;\frac{\mu\,B_{0}}{m}},
\label{eq:omega_par}
\end{equation} 
and the connection length $R_{\|} \equiv q(\psi)\,R$. Using these definitions and Eq.~\eqref{eq:ppar_equat}, the equatorial poloidal angular velocity is defined as
\begin{equation}
\dot{\vartheta}_{\rm e} \;\equiv\; 2\,\sqrt{\kappa}\;\omega_{\|}.
\label{eq:theta_dot_e}
\end{equation}
Note that the parallel frequency \eqref{eq:omega_par} is small compared to the gyrofrequency $\Omega_{0}$ (evaluated at the magnetic axis):
\begin{equation}
\frac{\omega_{\|}}{\Omega_{0}} \;=\; \frac{1}{R_{\|}}\;\sqrt{\epsilon\;\frac{\mu\,B_{0}}{m\Omega_{0}^{2}}} \;\equiv\; 
\sqrt{\frac{\epsilon}{2}}\;\frac{\rho_{0}}{R_{\|}} \;\ll\; 1,
\label{eq:omega_ratio}
\end{equation}
where $\rho_{0} \equiv \sqrt{2\mu B_{0}/(m\Omega_{0}^{2})}$ denotes the gyroradius of a guiding-center (with magnetic moment $\mu$) evaluated at the magnetic axis.

The condition $\kappa < 1$ in Eq.~\eqref{eq:ppar_tok} implies that a trapped particle bounces back and forth between the bounce poloidal angles $\pm\,
\vartheta_{\rm b}$, where
\begin{equation}
\vartheta_{\rm b} \;\equiv\; 2\;\arcsin\sqrt{\kappa}, 
\label{eq:thetab_def}
\end{equation}
while a passing particle is in transit for $\kappa > 1$ in Eq.~\eqref{eq:ppar_tok}, with the minimum parallel momentum $p_{\|{\rm min}} = 
p_{\|{\rm e}}\,\sqrt{1 - \kappa^{-1}} < p_{\|{\rm e}}$ attained at $\vartheta = \pi$. Using the definition \eqref{eq:kappa_def}, the bounce-transit boundary $\kappa = 1$ yields the trapping condition
\begin{equation}
\xi_{\rm e}^{2} \;<\; 2\epsilon/(1 + \epsilon) \;\simeq\; 2\,\epsilon
\label{eq:bt_boundary}
\end{equation}
in the simple tokamak magnetic field \eqref{eq:B_tok}.

\section{\label{sec:action}Bounce and Transit Actions}

At the lowest order on the bounce-transit time scale, the magnetic flux $\psi$ is frozen and thus the minor radius $r = \epsilon R$ and the connection length $R_{\|}$ are constant. Hence, the parallel momentum \eqref{eq:ppar_tok} depends only on the poloidal angle $\vartheta$ as a guiding-center moves along a magnetic-field line $s(\vartheta)$ (projected onto the poloidal plane at fixed toroidal angle $\phi$). 

\subsection{Bounce action and bounce frequency for trapped particles}

By using the infinitesimal parallel-length element $ds \simeq R_{\|}\,d\vartheta$, we easily obtain the following expression for the bounce action for trapped particles $(\kappa < 1$):
\begin{eqnarray}
J_{\rm b} & \equiv & \frac{1}{2\pi}\;\oint\;p_{\|}\,ds \nonumber \\
 & \simeq & p_{\|{\rm e}}\,R_{\|}\;\int_{-\,\vartheta_{\rm b}}^{\vartheta_{\rm b}}\;
\sqrt{1 \;-\; \kappa^{-1}\;\sin^{2}\frac{\vartheta}{2}}\;\frac{d\vartheta}{\pi}.
\label{eq:Jb_bounce}
\end{eqnarray}
After making the substitution $\sin\vartheta/2 = \sqrt{\kappa}\,\sin\varphi$ in Eq.~\eqref{eq:Jb_bounce}, we obtain \cite{Wang_Hahm}
\begin{eqnarray}
J_{\rm b} & = & p_{\|{\rm e}}R_{\|}\;\frac{4\sqrt{\kappa}}{\pi}\;\int_{0}^{\pi/2}\;\frac{\cos^{2}\varphi\;d\varphi}{
\sqrt{1 - \kappa\;\sin^{2}\varphi}} \nonumber \\
 & \equiv & m\,R_{\|}^{2}\,\omega_{\|}\; \left\{ \frac{8}{\pi}\left[ {\sf E}(\kappa) \;-\frac{}{} 
\left(1 -\frac{}{} \kappa \right)\;{\sf K}(\kappa) \right]\right\},
\label{eq:Jb_elliptic}
\end{eqnarray}
where we wrote $\cos^{2}\varphi = \kappa^{-1}\,[(1 - \kappa\,\sin^{2}\varphi) - (1 - \kappa)]$ while ${\sf K}(\kappa)$ and ${\sf E}(\kappa)$ denote the complete elliptic integrals of the first and second kind, respectively \cite{HMF_Jacobi,Brizard_elliptic}. 

By using standard properties of these complete elliptic integrals \cite{HMF_Jacobi}, the bounce frequency for trapped particles $\omega_{\rm b} \equiv 
(\partial J_{\rm b}/\partial{\cal E})^{-1}$ is expressed as
\begin{equation}
\omega_{\rm b} \;=\; \frac{\pi\,\omega_{\|}}{2\,{\sf K}(\kappa)},
\label{eq:omegab_bounce}
\end{equation}
where we used $\partial\kappa/\partial{\cal E} = \kappa/({\cal E} - \mu\,B_{\rm e})$ and the relation \cite{HMF_Jacobi}
\begin{equation}
\frac{d}{d\kappa}\left[{\sf E}(\kappa) \;-\frac{}{} (1 - \kappa)\;{\sf K}(\kappa)\right] \;=\; 
\frac{1}{2}\;{\sf K}(\kappa).
\label{eq:elliptic_int}
\end{equation}
In the deeply-trapped approximation, where the equatorial pitch-angle coordinate is $\xi_{\rm e}^{2} \ll 2\,\epsilon$ (so that $\kappa \ll 1$), 
we use ${\sf K}(\kappa) \simeq \pi/2$ and ${\sf E}(\kappa) - (1 - \kappa)\,{\sf K}(\kappa) \simeq \pi\,\kappa/4$, and 
Eqs.~\eqref{eq:Jb_elliptic}-\eqref{eq:omegab_bounce} become $J_{{\rm b}} \simeq {\cal E}\,\xi_{\rm e}^{2}/\omega_{\|0}$ and $\omega_{{\rm b}} \simeq \omega_{\|0}$, where $\omega_{\|0} = R_{\|}^{-1}\sqrt{\epsilon\;{\cal E}/m}$.
 
\begin{figure}
\epsfysize=2in
\epsfbox{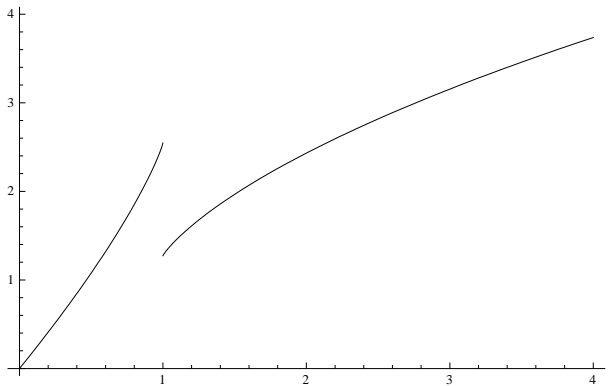}
\caption{Normalized action $\ov{J}(\kappa) \equiv J/(mR_{\|}^{2}\omega_{\|})$, given by Eq.~\eqref{eq:J_bar}, versus the bounce-transit parameter 
$\kappa$. The normalized bounce action \eqref{eq:Jb_bar} is shown for $\kappa < 1$ while the transit action \eqref{eq:Jt_bar} is shown for $\kappa > 1$.}
\label{fig:action} 
\end{figure}

\subsection{Transit action and transit frequency for passing particles}

For a passing particle $(\kappa > 1$), the transit action is
\begin{eqnarray}
J_{\rm t} & \equiv & p_{\|{\rm e}}R_{\|}\;\int_{-\pi}^{\pi}\;\sqrt{1 - \kappa^{-1}\;\sin^{2}\frac{\vartheta}{2}}\;
\frac{d\vartheta}{2\pi} \nonumber \\
 & = & m\,R_{\|}^{2}\,\omega_{\|}\;\left[ \frac{4\sqrt{\kappa}}{\pi}\;{\sf E}(\kappa^{-1}) \right].
\label{eq:J_transit}
\end{eqnarray}
The corresponding transit frequency $\omega_{\rm t} \equiv (\partial J_{\rm t}/\partial{\cal E})^{-1}$ is expressed as
\begin{equation}
\omega_{\rm t} \;=\; \omega_{\|}\;\frac{\pi\,\sqrt{\kappa}}{{\sf K}(\kappa^{-1})} \;\equiv\; \frac{\pi\,\omega_{\|}}{{\sf K}(\kappa)},
\label{eq:omega_transit}
\end{equation}
where we used the relation \cite{HMF_Jacobi}
\[ \frac{d}{d\kappa} \left[ \sqrt{\kappa}\;{\sf E}(\kappa^{-1}) \right] \;=\; \frac{{\sf K}(\kappa^{-1})}{2\,\sqrt{\kappa}} \;\equiv\; \frac{1}{2}\;
{\sf K}(\kappa). \] 
In the large-$\kappa$ (energetic-passing) limit, we recover $J_{{\rm t}} \simeq p_{\|{\rm e}}R_{\|}$ and $\omega_{{\rm t}} \simeq 
p_{\|{\rm e}}/(mR_{\|})$ from Eqs.~\eqref{eq:J_transit}-\eqref{eq:omega_transit}, which both scale as $\sqrt{\kappa}$ for $\kappa \gg 1$.

\begin{figure}
\epsfysize=2in
\epsfbox{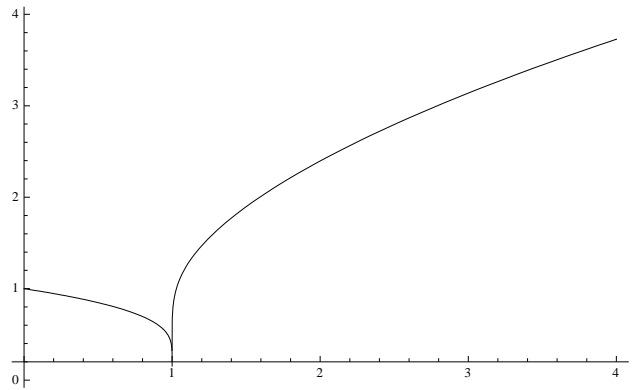}
\caption{Normalized frequency $\ov{\omega} \equiv \omega/\omega_{\|}$ versus the bounce-transit parameter $\kappa$. The bounce frequency 
\eqref{eq:omegab_bounce} is shown for $\kappa < 1$ while the transit frequency \eqref{eq:omega_transit} is shown for $\kappa > 1$.}
\label{fig:omega} 
\end{figure}

\subsection{Unified representation for action integrals}

We note that the bounce action \eqref{eq:Jb_elliptic} and the transit action \eqref{eq:J_transit} can both be combined into a single expression for the
normalized action $\ov{J}(\kappa) \equiv J/(mR_{\|}^{2}\omega_{\|})$ for arbitrary bounce-transit parameter $\kappa$:
\begin{equation}
\ov{J}(\kappa) \;\equiv\; \frac{8}{\pi}\,\alpha(\kappa)\,\kappa\;\int_{0}^{{\sf K}(\kappa)}\;{\rm cn}^{2}(u\,|\,\kappa)\; du,
\label{eq:J_bar}
\end{equation}
where $\alpha(\kappa) = 1\;(\kappa < 1)$ or $\alpha(\kappa) = 1/2\;(\kappa > 1)$. Note that the factor $\alpha = 1/2$ yields a consistent relation between the poloidal angle $\vartheta$ and the transit angle $\zeta_{\rm t}$ for passing particles [see Eq.~\eqref{eq:zeta_transit}]. 

For $\kappa < 1$, we use the definition \cite{HMF_Jacobi} for the elliptic integral
\begin{equation}
{\sf E}(\kappa) \;\equiv\; \int_{0}^{{\sf K}(\kappa)}\;{\rm dn}^{2}(u\,|\,\kappa)\; du
\label{eq:E_complete}
\end{equation}
and the identities
\begin{equation} 
{\rm cn}^{2}(u|\kappa) \;+\; {\rm sn}^{2}(u|\kappa) \;\equiv\; 1 \;\equiv\; {\rm dn}^{2}(u|\kappa) \;+\; \kappa\;{\rm sn}^{2}(u|\kappa),
\label{eq:cnsndn_rel}
\end{equation}
to obtain 
\begin{eqnarray} 
\ov{J}_{\rm b}(\kappa) & \equiv & \frac{8}{\pi}\;\int_{0}^{{\sf K}(\kappa)}\;\left[\kappa \;-\; 1 \;+\; {\rm dn}^{2}(u\,|\,\kappa)\right]\; du \nonumber \\
 & = & \frac{8}{\pi} \left[ {\sf E}(\kappa) \;-\frac{}{} (1 - \kappa)\,{\sf K}(\kappa) \right],
\label{eq:Jb_bar}
\end{eqnarray}
so that $\ov{J}_{\rm b}(\kappa = 1) = 8/\pi$. For $\kappa > 1$, we use the conversion relation \cite{HMF_elliptic,HMF_Jacobi} ${\rm cn}(u|\kappa)
\equiv {\rm dn}(\sqrt{\kappa}\,u|\kappa^{-1})$ associated with $\kappa > 1 \rightarrow \kappa^{-1} < 1$. By using the change of variable $v = 
\sqrt{\kappa}\,u$, Eq.~\eqref{eq:J_bar} becomes
\begin{eqnarray}
\ov{J}_{\rm t}(\kappa) & \equiv & \frac{8\,\sqrt{\kappa}}{2\pi}\;\int_{0}^{{\sf K}(\kappa^{-1})}\;{\rm dn}^{2}(v\,|\,\kappa^{-1})\; dv \nonumber \\ 
 & = & \frac{4\,\sqrt{\kappa}}{\pi}\;{\sf E}(\kappa^{-1}),
\label{eq:Jt_bar}
\end{eqnarray}
so that $\ov{J}_{\rm t}(\kappa = 1) = 4/\pi$. 

Figures \ref{fig:action} and \ref{fig:omega} show the normalized action $\ov{J} \equiv J/(mR_{\|}^{2}\omega_{\|})$, defined by Eq.~\eqref{eq:J_bar}, and the normalized frequency $\ov{\omega} \equiv \omega/\omega_{\|}$ versus the bounce-transit parameter $\kappa$. Figure \ref{fig:action} shows the discontinuity of the bounce and transit actions at $\kappa= 1$. This discontinuity is related to the fact the transit period for a marginally-passing particle is associated with the motion from $\vartheta = -\,\pi$ to $+\,\pi$, while the bounce period for a marginally-trapped particle is associated with the motion from $\vartheta = -\,\pi$ to $+\,\pi$ and then back to $\vartheta = -\,\pi$ (i.e., doubling the path taken to complete a bounce period). The bounce frequency \eqref{eq:omegab_bounce} and the transit frequency \eqref{eq:omega_transit} both vanish (i.e., the corresponding periods become infinite) at the bounce-transit boundary $\kappa = 1$ since ${\sf K}(x) \rightarrow \infty$ as $x \rightarrow 1$.

\section{\label{sec:angle}Bounce and Transit Angles}

The angles canonically conjugate to the bounce action $J_{\rm b}$ and the transit action $J_{\rm t}$ are the bounce angle $0 \leq \zeta_{\rm b} \leq 2\pi$ and the transit angle $-\,\pi \leq \zeta_{\rm t} \leq \pi$, respectively. These angles have been expressed previously in terms of the incomplete elliptic integrals, which often prevented immediate analytical applications. In the present Section, we use the Jacobi elliptic functions to give compact expressions of the poloidal angle $\vartheta$ in terms of the bounce or transit angles.

\subsection{Bounce angle for trapped particles}

The bounce angle $\zeta_{\rm b}$ for trapped particles $(\kappa < 1)$ is defined as
\begin{eqnarray} 
\zeta_{\rm b} & \equiv & \pi \;+\; \frac{\omega_{\rm b}}{2\sqrt{\kappa}\,\omega_{\|}}\;\int_{-\vartheta_{\rm b}}^{\vartheta}\;\frac{d\vartheta'}{
\sqrt{1 - \kappa^{-1}\sin^{2}\vartheta'/2}} \nonumber \\
 & = & \pi \;+\; \frac{\pi}{2\,{\sf K}(\kappa)}\;\int^{\pi/2}_{-\,\Theta}\;\frac{d\varphi}{\sqrt{1 - \kappa\;
\sin^{2}\varphi}} \nonumber \\
 & \equiv & \frac{3\pi}{2} \;+\; \frac{\pi}{2\,{\sf K}(\kappa)}\;{\rm sn}^{-1}\left( \sin\Theta \;|\frac{}{} \kappa\right),
\label{eq:zeta_bounce}
\end{eqnarray}
where $\sin\Theta \equiv \sin(\vartheta/2)/\sqrt{\kappa}$ and we defined $\zeta_{\rm b} \equiv \pi$ at $\vartheta = -\,\vartheta_{\rm b}$ (i.e., at 
$\Theta = -\,\pi/2$). Equation \eqref{eq:zeta_bounce} can be inverted to give
\begin{eqnarray}
\sin\frac{\vartheta}{2} & = & \sqrt{\kappa}\;{\rm sn}\left( \frac{2{\sf K}(\kappa)}{\pi}\,\zeta_{\rm b} \;-\; 3\,{\sf K}(\kappa) \;\left|\frac{}{} \kappa
\right. \right) \nonumber \\
 & \equiv & \sqrt{\kappa}\;{\rm cd}\left(2{\sf K}(\kappa)\,\zeta_{\rm b}/\pi\;\left|\frac{}{} \kappa\right. \right),
\label{eq:sin_theta_bounce}
\end{eqnarray}
where we used the standard notation ${\rm pq} \equiv {\rm pn}/{\rm qn}$ (with p and q either c, s, or d) and the identities \cite{HMF_Jacobi} 
${\rm sn}(u - 3\,{\sf K}) \equiv {\rm sn}(u + {\sf K}) \equiv {\rm cd}(u)$, with $u \equiv 2{\sf K}(\kappa)\,\zeta_{\rm b}/\pi$ and we used the 
$4\,{\sf K}$-periodicity of ${\rm sn}$.

\begin{figure}
\epsfysize=2in
\epsfbox{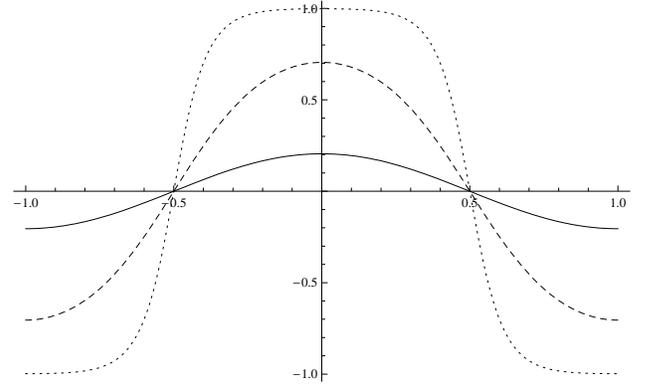}
\caption{Normalized poloidal angle $\vartheta/\pi$ versus normalized bounce angle $\zeta_{\rm b}/\pi$ for $\kappa = 0.1$ (solid curve), 0.8 (dashed curve), and 0.999 (dotted curve). In the deeply-trapped limit $(\kappa \ll 1)$, we find $\vartheta \simeq \vartheta_{\rm b}\;\cos\zeta_{\rm b}$ and 
$\vartheta \equiv 0$ at $\zeta_{\rm b} = \pi/2$ for all values of $\kappa < 1$.}
\label{fig:theta_bounce} 
\end{figure}

Equation \eqref{eq:sin_theta_bounce} yields the poloidal angular velocity
\begin{eqnarray} 
\dot{\vartheta} & = & -\;\dot{\vartheta}_{\rm e}\;\sqrt{1 - \kappa}\;{\rm sd}\left(2{\sf K}(\kappa)\,\zeta_{\rm b}/\pi\;\left|\frac{}{} \kappa
\right. \right) \nonumber \\ 
 & \equiv & -\;\dot{\vartheta}_{\rm e}\;\sqrt{1 \;-\; \kappa^{-1}\;\sin^{2}\frac{\vartheta}{2}},
\label{theta_dot_b}
\end{eqnarray}
where $\dot{\vartheta}_{\rm e}$ is defined in Eq.~\eqref{eq:theta_dot_e} and we used the relations \eqref{eq:cnsndn_rel}. Figure \ref{fig:theta_bounce} shows the normalized poloidal angle $\vartheta/\pi$ versus the normalized bounce angle $\zeta_{\rm b}/\pi$ for various values of $\kappa < 1$. In the standard deeply-trapped approximation $(\kappa \ll 1)$ \cite{Wang_Hahm} $\vartheta \simeq \vartheta_{\rm b}\;\cos\zeta_{\rm b}$ is shown by the solid curve. 

\subsection{Transit angle for passing particles}

\begin{figure}
\epsfysize=2in
\epsfbox{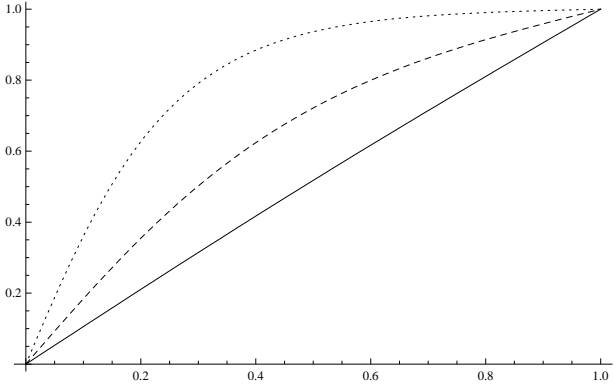}
\caption{Normalized poloidal angle $\vartheta/\pi$ versus normalized transit angle $\zeta_{\rm t}/\pi$ for $\kappa = 5$ (solid curve), 1.05 (dashed curve), and 1.0001 (dotted curve). In the energetic-passing limit $(\kappa^{-1} \ll 1)$, we find $\vartheta \simeq \zeta_{\rm t}$ (solid curve).}
\label{fig:theta_transit} 
\end{figure}

Next, the transit angle $\zeta_{\rm t}$ for passing particles $(\kappa > 1)$ is defined as
\begin{eqnarray}
\zeta_{\rm t} & \equiv & \frac{\omega_{\rm t}}{2\sqrt{\kappa}\,\omega_{\|}}\;\int_{0}^{\vartheta}\; \frac{d\vartheta^{\prime}}{
\sqrt{1 - \kappa^{-1}\;\sin^{2}\vartheta^{\prime}/2}} \nonumber \\
 & = & \frac{\pi}{{\sf K}(\kappa^{-1})}\;\int_{0}^{\vartheta/2}\;\frac{d\varphi}{\sqrt{1 - \kappa^{-1}\sin^{2}\varphi}} \nonumber \\
 & \equiv & \frac{\pi}{{\sf K}(\kappa^{-1})}\;{\rm sn}^{-1}\left( \sin\frac{\vartheta}{2}\;\left|\; \kappa^{-1}\frac{}{}\right.\right),
\label{eq:zeta_transit}
\end{eqnarray}
which can be easily inverted to give
\begin{eqnarray}
\sin\frac{\vartheta}{2} & = & {\rm sn}\left( {\sf K}(\kappa^{-1})\;\frac{\zeta_{\rm t}}{\pi} \;\left|\; \kappa^{-1}\frac{}{}\right. \right) \nonumber \\
 & \equiv & \sqrt{\kappa}\;{\rm sn}\left( {\sf K}(\kappa)\;\frac{\zeta_{\rm t}}{\pi} \;\left|\; \kappa\frac{}{}\right. \right),
\label{eq:sin_theta_transit}
\end{eqnarray}
with $\vartheta = \pi$ when $\zeta_{\rm t} = \pi$ (for all values of $\kappa > 1$) and we used ${\rm sn}(\sqrt{\kappa}\,u|\kappa^{-1}) \equiv 
\sqrt{\kappa}\;{\rm sn}(u|\kappa)$ and ${\sf K}(\kappa^{-1}) \equiv \sqrt{\kappa}\,{\sf K}(\kappa)$. We note that the definition \eqref{eq:zeta_transit} yields the correct poloidal angular velocity
\begin{equation} 
\dot{\vartheta} \;=\; \dot{\vartheta}_{\rm e}\;{\rm cn}\left({\sf K}(\kappa)\,\zeta_{\rm t}/\pi\;\left|\frac{}{} \kappa\right. \right) \;\equiv\; 
\dot{\vartheta}_{\rm e}\;\sqrt{1 \;-\; \kappa^{-1}\;\sin^{2}\frac{\vartheta}{2}},
\label{theta_dot_t}
\end{equation} 
which is consistent with the definition of the transit action \eqref{eq:J_transit}. Figure \ref{fig:theta_transit} shows the normalized poloidal angle 
$\vartheta/\pi$ versus normalized transit angle $\zeta_{\rm t}/\pi$ for various values of $\kappa^{-1} < 1$. Note that in the energetic-passing limit 
$\kappa^{-1} \ll 1$, we find \cite{Wang_Hahm} $\vartheta \simeq \zeta_{\rm t}$ (shown by a solid line in Fig.~\ref{fig:theta_transit}).

\section{\label{sec:canonical}Canonical Parallel Coordinates}

In this Section, we derive suitable expressions for the canonical parallel coordinates $(s, p_{\|})$, which are required to satisfy (with $ds \simeq 
R_{\|}\,d\vartheta$) the canonical condition
\begin{eqnarray}
\{ s,\; p_{\|} \} & \equiv & \pd{s}{\zeta}\;\pd{p_{\|}}{J} \;-\; \pd{s}{J}\;\pd{p_{\|}}{\zeta} \label{eq:sp_canonical} \\
 & = & \frac{({\cal E} - \mu\,B_{\rm e})}{p_{\|{\rm e}} \sqrt{\kappa}} \left( \pd{\vartheta}{\chi}\;\pd{p_{\|}}{\cal E} \;-\; \pd{\vartheta}{\cal E}\;\pd{p_{\|}}{\chi} \right) \;=\; 1
\nonumber
\end{eqnarray}
for both trapped and transit particles, where $\zeta \equiv 2\pi\,\chi\,[\alpha(\kappa)/{\sf K}(\kappa)]$ for trapped-particle $(\alpha = 1)$ and passing-particle $(\alpha = 1/2)$ orbits. In Eq.~\eqref{eq:sp_canonical}, we also made use of the identity $R_{\|}\omega_{\|} \equiv ({\cal E} - \mu\,
B_{\rm e})/(p_{\|{\rm e}}\sqrt{\kappa})$, which follows from the definitions \eqref{eq:ppar_equat}-\eqref{eq:omega_par}.

\subsection{Parallel coordinates for trapped particles}

\begin{figure}
\epsfysize=2in
\epsfbox{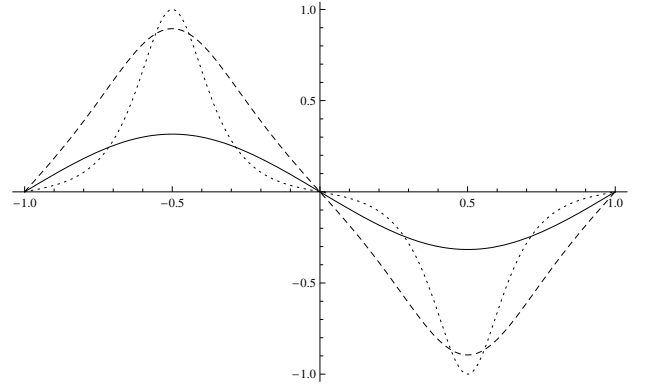}
\caption{Normalized parallel momentum $p_{\|}/(mR_{\|}\omega_{\|})$ versus the normalized bounce angle $\zeta_{\rm b}/\pi$ for $\kappa = 0.1$ (solid curve), 0.8 (dashed curve), and 0.999 (dotted curve). In the deeply-trapped limit $(\kappa \ll 1)$, $p_{\|} \simeq -\,p_{\|{\rm e}}\;\sin\zeta_{\rm b}$ (solid curve).}
\label{fig:ppar_bounce} 
\end{figure}

For a trapped particle, using Eq.~\eqref{eq:sin_theta_bounce}, the parallel momentum \eqref{eq:ppar_tok} is defined as
\begin{eqnarray}
p_{\|} & = & -\,p_{\|{\rm e}}\;\sqrt{1 \;-\; \kappa^{-1}\;\sin^{2}\vartheta/2} \nonumber \\
 & \equiv & -\;p_{\|{\rm e}}\,\sqrt{1 - \kappa}\;{\rm sd}\left(\chi_{\rm b}\;\left|\frac{}{} \kappa\right. \right).
\label{eq:ppar_bounce} 
\end{eqnarray}
Figure \ref{fig:ppar_bounce} shows the normalized parallel momentum $p_{\|}/(mR_{\|}\omega_{\|})$ versus the normalized bounce angle $\zeta_{\rm b}/\pi$ for various values of $\kappa < 1$, where the standard deeply-trapped approximation $p_{\|}(\zeta_{\rm b}) \simeq -\;p_{\|{\rm e}}\;\sin\zeta_{\rm b}$ is shown as a solid curve. 

Next, we verify that the parallel coordinates $\vartheta({\cal E}, \chi_{\rm b})$ and $p_{\|}({\cal E}, \chi_{\rm b})$, defined by Eqs.~\eqref{eq:sin_theta_bounce} and \eqref{eq:ppar_bounce}, satisfy the canonical condition 
\eqref{eq:sp_canonical}. First, using Eq.~\eqref{eq:sin_theta_bounce}, we obtain
\begin{equation}
\pd{\vartheta}{\chi_{\rm b}} \;=\; -\,2\;\sqrt{\kappa\,(1 - \kappa)}\;{\rm sd},
\label{eq:theta_psi_b}
\end{equation}
and
\begin{equation}
\pd{\vartheta}{\cal E} \;=\; \frac{\sqrt{\kappa/(1 - \kappa)}}{({\cal E} - \mu\,B_{\rm e})} \left[ {\rm cn} \;+\; {\rm dn}\;\left(2\,\kappa\;
\pd{{\rm cd}}{\kappa} \right) \right],
\label{eq:theta_E_b}
\end{equation}
where the expression for $\partial{\rm cd}/\partial\kappa$ is not needed in what follows \cite{footnote}. Next, using Eq.~\eqref{eq:ppar_bounce}, we obtain
\begin{equation}
\pd{p_{\|}}{\chi_{\rm b}} \;=\; -\,p_{\|{\rm e}}\;\sqrt{1 - \kappa}\;{\rm cd}\;{\rm nd},
\label{eq:p_psi_b}
\end{equation}
and
\begin{equation}
\pd{p_{\|}}{\cal E} \;=\; \frac{-\,p_{\|{\rm e}}}{2\,({\cal E} - \mu\,B_{\rm e})} \left[ \frac{(1 - 2\kappa)}{\sqrt{1 - \kappa}}\;{\rm sd} \;+\;
2\,\kappa\,\sqrt{1 - \kappa}\;\pd{{\rm sd}}{\kappa} \right].
\label{eq:p_E_b}
\end{equation}
By combining Eqs.~\eqref{eq:theta_psi_b}-\eqref{eq:p_E_b} into Eq.~\eqref{eq:sp_canonical}, we obtain
\begin{equation}
\{ s,\; p_{\|}\}_{\rm b} \;=\; \pd{}{\kappa}\left[\;\kappa\;\left( (1 - \kappa)\,{\rm sd}^{2} \;+\frac{}{} {\rm cd}^{2}\right) \;\right] \;=\; 1,
\label{eq:canonical_bounce}
\end{equation}
where we used the identity $(1 - \kappa)\,{\rm sd}^{2} + {\rm cd}^{2} \equiv 1$ obtained from Eq.~\eqref{eq:cnsndn_rel}.

\subsection{Parallel coordinates for passing particles}

\begin{figure}
\epsfysize=2in
\epsfbox{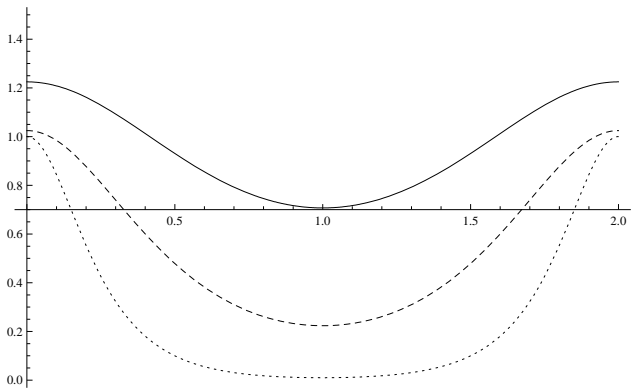}
\caption{Normalized parallel momentum $p_{\|}/(mR_{\|}\omega_{\|})$ versus the normalized transit angle $\zeta_{\rm t}/\pi$ for $\kappa = 1.5$ (solid curve), 1.05 (dashed curve), and 1.0001 (dotted curve). The modulation of the parallel momentum of a passing particle decreases as $\kappa$ increases.}
\label{fig:ppar_transit} 
\end{figure}

For a passing particle, using Eq.~\eqref{eq:sin_theta_transit}, the parallel momentum \eqref{eq:ppar_tok} is
\begin{eqnarray}
p_{\|} & = & p_{\|{\rm e}}\; {\rm dn}\left( {\sf K}(\kappa^{-1})\;\frac{\zeta_{\rm t}}{\pi} \;\left|\frac{}{} \kappa^{-1}\right. \right) \nonumber \\
 & \equiv & p_{\|{\rm e}}\;{\rm cn}\left( {\sf K}(\kappa)\;\frac{\zeta_{\rm t}}{\pi} \;\left|\frac{}{} \kappa\right. \right),
\label{eq:ppar_transit}
\end{eqnarray}
where $p_{\|{\rm min}} \equiv p_{\|{\rm e}}\,\sqrt{1 - \kappa^{-1}}$ when $\zeta_{\rm t} = \pi$. The modulation in parallel momentum 
\begin{eqnarray}
\Delta p_{\|} & = & p_{\|{\rm e}}\,\left( 1 \;-\; \sqrt{1 - \kappa^{-1}}\right) \nonumber \\ 
 & = & 2\,m\,R_{\|}\omega_{\|} \left( \sqrt{\kappa} \;-\frac{}{}\sqrt{\kappa - 1} \right)
\label{eq:Delta_ppar}
\end{eqnarray}
decreases with increasing values of $\kappa$. Figure \ref{fig:ppar_transit} shows the normalized parallel momentum $p_{\|}/(mR_{\|}
\omega_{\|})$ versus the normalized transit angle $\zeta_{\rm t}/\pi$ for various values of $\kappa > 1$. 

We now verify that the parallel coordinates $\vartheta({\cal E}, \chi_{\rm t})$ and $p_{\|}({\cal E}, \chi_{\rm t})$, defined by 
Eqs.~\eqref{eq:sin_theta_transit} and \eqref{eq:ppar_transit}, satisfy the canonical condition 
\eqref{eq:sp_canonical}. First, using Eq.~\eqref{eq:sin_theta_transit}, we obtain
\begin{equation}
\pd{\vartheta}{\chi_{\rm t}} \;=\; 2\,\sqrt{\kappa}\;{\rm cn},
\label{eq:theta_psi_t}
\end{equation}
and
\begin{equation}
\pd{\vartheta}{\cal E} \;=\; \frac{\sqrt{\kappa}}{({\cal E} - \mu\,B_{\rm e})} \left[\; {\rm sd} \;+\; {\rm nd} \left( 2\,\kappa\;\pd{{\rm sn}}{\kappa} \right) \right].
\label{eq:theta_E_t}
\end{equation}
Next, using Eq.~\eqref{eq:ppar_transit}, we obtain
\begin{equation}
\pd{p_{\|}}{\chi_{\rm t}} \;=\; -\,p_{\|{\rm e}}\;{\rm dn}\;{\rm sn},
\label{eq:p_psi_t}
\end{equation}
and
\begin{equation}
\pd{p_{\|}}{\cal E} \;=\; \frac{p_{\|{\rm e}}}{2\,({\cal E} - \mu\,B_{\rm e})} \left( {\rm cn} \;+\; 2\kappa\;\pd{{\rm cn}}{\kappa} \right).
\label{eq:p_E_t}
\end{equation}
By combining Eqs.~\eqref{eq:theta_psi_t}-\eqref{eq:p_E_t} into Eq.~\eqref{eq:sp_canonical}, we obtain 
\begin{equation}
\{ s,\; p_{\|}\}_{\rm t} \;=\; \pd{}{\kappa} \left[\;\kappa\;\left({\rm sn}^{2} \;+\frac{}{} {\rm cn}^{2}\right)\;\right] \;=\; 1,
\label{eq:canonical_transit}
\end{equation}
which follows from Eq.~\eqref{eq:cnsndn_rel}.

\subsection{Phase portrait}

We have demonstrated in Eqs.~\eqref{eq:canonical_bounce} and \eqref{eq:canonical_transit} that the parallel coordinates $(s, p_{\|})$, expressed in terms of Jacobi elliptic functions, are valid canonical coordinates for the trapped-particle and passing-particle guiding-center orbits. 

\begin{figure}
\epsfysize=1.5in
\epsfbox{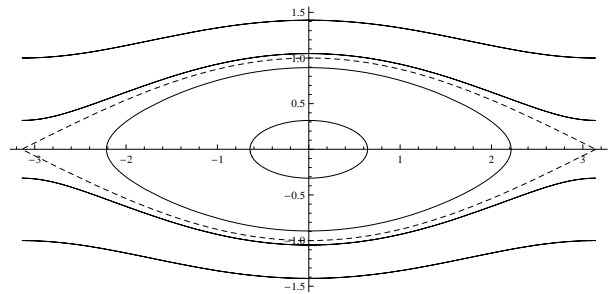}
\caption{Phase portrait in $(\vartheta,\dot{\vartheta})$-plane of guiding-center trapped and transit orbits. The separatrix (dashed curves) separates trapped-particle orbits (inside) from the passing-particle orbits (outside).}
\label{fig:tok_elliptic} 
\end{figure}

Figure \ref{fig:tok_elliptic} shows the phase portrait of the guiding-center trapped-particle and passing-particle orbits in the $(\vartheta,
\dot{\vartheta})$-plane, which combines expressions for the the poloidal angle $\vartheta$ and the normalized poloidal angular velocity 
$\dot{\vartheta}/\dot{\vartheta}_{\rm e}$ for trapped and passing particles. Here, we easily recognize the well-known separatrix structure of the standard pendulum problem, which separates trapped-particle orbits (inside) from passing-particle orbits (outside). In Sec.~\ref{sec:toroidal}, we will project the trapped-particle and passing-particle guiding-center orbits onto the poloidal plane (at a constant toroidal angle $\phi$).

\section{\label{sec:toroidal}Toroidal Guiding-center Dynamics}

In this Section, we discuss the guiding-center dynamics in the toroidal direction (i.e., the direction of axisymmetry) and its implications on the canonically-conjugate drift action. So far, the toroidal angle $\phi$ was considered fixed, which enabled a study of the trapped and passing guiding-center orbits projected onto the poloidal plane, and the magnetic flux $\psi$ was assumed to be frozen on the bounce-transit time scale.

The axisymmetry of the tokamak magnetic geometry implies that the toroidal canonical momentum (or drift action) 
\begin{equation}
P_{\phi} \;=\; -\;\frac{e}{c}\;\psi \;+\; p_{\|}\;\frac{B_{\phi}}{B} \;\equiv\; -\,\frac{e}{c}\;\psi_{*} 
\label{eq:P_phi_def}
\end{equation}
is conserved, where $B_{\phi} \simeq B_{0}\,R$ is approximately constant. 

\subsection{Trapped-particle guiding-center orbits}

For a trapped-particle guiding-center orbit, this conservation law is expressed as $\psi_{*} \equiv \psi_{0}$ in terms of the magnetic flux $\psi_{0}$ at the bounce points, so that Eq.~\eqref{eq:P_phi_def} becomes
\begin{equation}
\psi \;-\; \rho_{\|}\,B_{\phi} \;\equiv\; \psi_{0},
\label{eq:psi_star}
\end{equation}
where $\rho_{\|} \equiv p_{\|}/(m\Omega)$. By substituting the Taylor approximation $\psi \simeq \psi_{0} + 
\Delta\,r\;(\epsilon_{0}\,B_{\phi}/q_{0})$, the normalized radial deviation $\Delta \epsilon_{\rm b} \equiv \epsilon - \epsilon_{0}$ (from the magnetic surface $\psi \equiv \psi_{0}$) is expressed as
\begin{equation}
\Delta\epsilon_{\rm b} \;\simeq\; \frac{q_{0}}{\epsilon_{0}}\;\frac{\rho_{\|}}{R} \;=\; -\,2\,\delta\; \sqrt{\kappa\,(1 - \kappa)}\;
{\rm sd}\left(2{\sf K}(\kappa)\,\zeta_{\rm b}/\pi\;\left|\frac{}{} \kappa\right. \right),
\label{eq:Delta_epsilon_b}
\end{equation}
where, using Eq.~\eqref{eq:omega_ratio}, we have defined the trapped-orbit parameter
\begin{equation}
\delta \;\equiv\; \frac{q_{0}^{2}\,\omega_{\|}}{\epsilon_{0}\,\Omega_{0}} \;=\; \frac{q_{0}}{\sqrt{2\epsilon_{0}}}\;\frac{\rho_{0}}{R}.
\label{eq:delta_b}
\end{equation} 
The maximum radial deviation \eqref{eq:Delta_epsilon_b} is $(\Delta\epsilon_{\rm b})_{\rm max} = 2\delta\,\sqrt{\kappa} \leq 2\delta$. 

\begin{figure}
\epsfysize=2.5in
\epsfbox{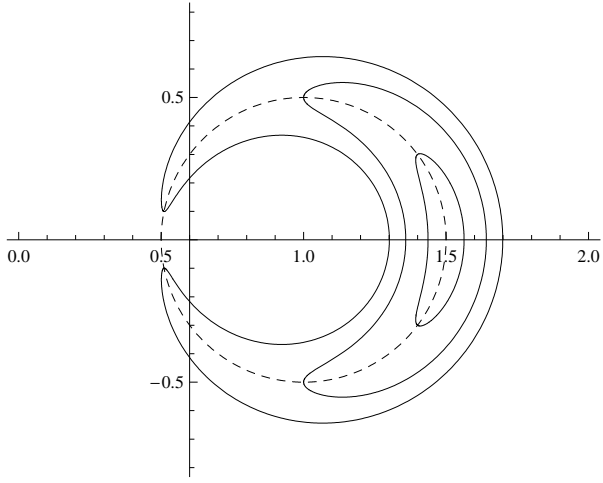}
\caption{Trapped (banana) guiding-center orbits in the normalized poloidal $(\ov{x},\ov{z})$-plane for $\kappa = (0.1, 0.5, 0.99)$, with the invariant 
$\psi^{*} = \psi_{0}$ chosen as the magnetic flux $\psi_{0}$ of the bounce points (shown as a dashed circle).}
\label{fig:banana_orbit} 
\end{figure}

A trapped-particle guiding-center orbit can be projected onto the normalized poloidal plane $(\ov{x} \equiv x/R, \ov{z} \equiv z/R)$ and represented as
\begin{equation}
\left.\begin{array}{rcl}
\ov{x}(\zeta_{\rm b},\kappa) & = & 1 \;+\; \left(\epsilon_{0} \;+\frac{}{} \Delta\epsilon_{\rm b} \right) \cos\vartheta(\zeta_{\rm b},\kappa) \\
 & & \\
\ov{z}(\zeta_{\rm b},\kappa) & = & \left(\epsilon_{0} \;+\frac{}{} \Delta\epsilon_{\rm b} \right) \sin\vartheta(\zeta_{\rm b},\kappa)
\end{array} \right\}.
\label{eq:banana_eq}
\end{equation}
Figure \ref{fig:banana_orbit} shows the trapped (banana) guiding-center orbits for $(\epsilon_{0},\delta) = (0.5,0.1)$ for various values of 
$\kappa < 1$, with the magnetic surface $\psi = \psi_{0}$ shown as a dashed circle. We note that, in the simple orbit topology considered here, a trapped-particle orbit does not enclose the magnetic axis \cite{Eriksson_Porcelli}.

\subsection{Passing-particle guiding-center orbits}

For a passing-particle guiding-center orbit, the conservation law \eqref{eq:psi_star} may be expressed in terms of the magnetic flux $\psi_{*} \equiv 
\psi_{\rm t}$ at the inside equatorial midplane (at $\vartheta = \pi$). By using the Taylor approximation $\psi \simeq 
\psi_{\rm t} + \Delta\,r\;(\epsilon_{\rm t}\,B_{\phi}/q_{\rm t})$, the normalized radial deviation $\Delta \epsilon_{\rm t} \equiv \epsilon - 
\epsilon_{\rm t}$ (from the magnetic surface $\psi \equiv \psi_{\rm t}$) is expressed as
\begin{equation}
\Delta\epsilon_{\rm t} \;\equiv\; 2\,\delta\;\left[\; \sqrt{\kappa}\,{\rm dn}\left({\sf K}(\kappa^{-1})\,\frac{\zeta_{\rm t}}{\pi}\;\left|\frac{}{} 
\kappa^{-1}\right. \right) \;-\; \sqrt{\kappa - 1} \;\right],
\label{eq:Delta_epsilon_t}
\end{equation}
where orbit parameter \eqref{eq:delta_b} is now evaluated at $\psi_{\rm t}$ (note that $\Delta\epsilon_{\rm t}$ vanishes on the inside equatorial plane where $p_{\|} = p_{\|{\rm min}} = p_{\|{\rm e}}\,\sqrt{1 - \kappa^{-1}}$). The maximum radial deviation \eqref{eq:Delta_epsilon_t} is 
$(\Delta\epsilon_{\rm t})_{\rm max} = 2\delta\,(\sqrt{\kappa} - \sqrt{\kappa - 1})$, i.e., the largest radial deviation for a passing-particle orbit occurs for marginally-passing particles $(\kappa - 1 \ll 1)$.

\begin{figure}
\epsfysize=2.5in
\epsfbox{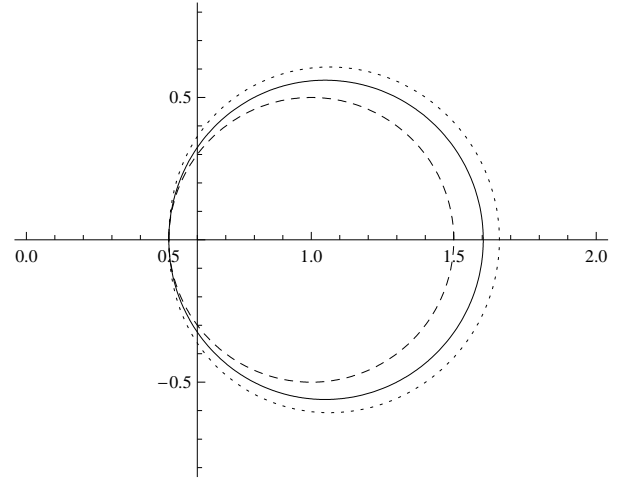}
\caption{Passing guiding-center orbits in the normalized poloidal $(x,z)$-plane, with the invariant $\psi^{*} = \psi_{\rm t}$ chosen as the magnetic flux $\psi_{\rm t}$ on the inside equatorial plane (shown as a dashed circle). The dotted and solid circles are passing guiding-center orbits with $\kappa = 1.05$ and $1.5$, respectively.}
\label{fig:passing_orbit} 
\end{figure}

A passing-particle guiding-center orbit can be projected onto the normalized poloidal plane and represented as
\begin{equation}
\left.\begin{array}{rcl}
\ov{x}(\zeta_{\rm t},\kappa) & = & 1 \;+\; \left(\epsilon_{\rm t} \;+\frac{}{} \Delta\epsilon_{\rm t} \right) \cos\vartheta(\zeta_{\rm t},\kappa) \\
 & & \\
\ov{z}(\zeta_{\rm t},\kappa) & = & \left(\epsilon_{\rm t} \;+\frac{}{} \Delta\epsilon_{\rm t} \right) \sin\vartheta(\zeta_{\rm t},\kappa)
\end{array} \right\}.
\label{eq:passing_eq}
\end{equation}
Figure \ref{fig:passing_orbit} shows the passing guiding-center orbits for $(\epsilon_{\rm t},\delta) = (0.5,0.1)$ for various values of 
$\kappa > 1$, with the magnetic surface $\psi = \psi_{\rm t}$ shown as a dashed circle.

\subsection{Bounce-averaged toroidal drift frequency}

\begin{figure}
\epsfysize=2in
\epsfbox{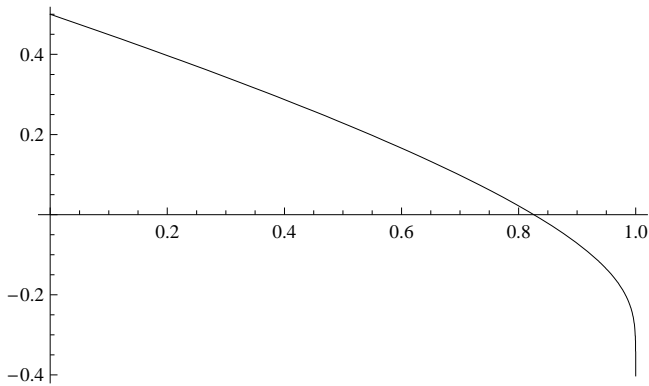}
\caption{Normalized drift frequency $[{\sf E}(\kappa)/{\sf K}(\kappa)] - 1/2$ as a function of $\kappa$. Note that the toroidal drift frequency 
\eqref{eq:omega_drift} changes sign for $\kappa > \kappa_{\rm r} \simeq 0.826$.}
\label{fig:drift} 
\end{figure}

The bounce-averaged toroidal drift frequency \cite{RBW}
\begin{equation}
\omega_{\rm d} \;\simeq\; \left\langle\dot{\phi} \;-\frac{}{} q\;\dot{\vartheta} \right\rangle \;\equiv\; -\;\frac{c\mu}{e}\;\left\langle\pd{B}{\psi}\right\rangle
\label{eq:omega_d}
\end{equation}
is approximated (ignoring magnetic shear) as
\begin{equation}
\omega_{\rm d} \;\simeq\; \frac{2\mu B_{0}}{m\Omega_{0}}\;\frac{q}{\epsilon\,R^{2}} \left( 
\frac{{\sf E}(\kappa)}{{\sf K}(\kappa)} \;-\; \frac{1}{2} \right),
\label{eq:omega_drift}
\end{equation}
where we used $\partial B/\partial\psi \simeq -\,(q/\epsilon R^{2})\,\cos\vartheta$. We note that the bounce-averaged toroidal drift frequency \eqref{eq:omega_drift} exhibits a drift reversal for $\kappa > \kappa_{\rm r} \simeq 0.826$, where ${\sf K}(\kappa_{\rm r}) \equiv 2\,{\sf E}(\kappa_{\rm r})$ (see Fig.~\ref{fig:drift}); a more complete expression for the bounce-averaged toroidal drift frequency can be found elsewhere \cite{Galeev_Sagdeev}. 

Lastly, the toroidal drift frequency \eqref{eq:omega_drift} can also be expressed in terms of a radial derivative of the bounce action as follows. First, we write 
\begin{eqnarray} 
\pd{J_{\rm b}}{\epsilon} & \simeq & \frac{4m}{\pi\,\epsilon}\,R_{\|}^{2}\omega_{\|}\left[ {\sf E} - 
\left(1 -\frac{}{} \kappa \right)\;{\sf K} + \left( \frac{1}{2} - \kappa\right) {\sf K} \right] \nonumber \\
 & = & 2\;
\frac{\mu B_{0}}{\omega_{\rm b}} \left( \frac{{\sf E}}{{\sf K}} \;-\; \frac{1}{2} \right),
\label{eq:J_epsilon}
\end{eqnarray}
where we used Eq.~\eqref{eq:elliptic_int} and $\partial\kappa/\partial\epsilon = (1/2 - \kappa)/\epsilon$. The toroidal drift frequency 
\eqref{eq:omega_drift} can thus be approximately expressed as
\begin{equation}
\omega_{\rm d} \;\simeq\; \frac{q\omega_{\rm b}}{\epsilon\,\Omega_{0}} \;\pd{}{\epsilon}\left(\frac{J_{\rm b}}{m\,R^{2}} \right).
\label{eq:omega_drift_epsilon}
\end{equation}
Hence, the toroidal drift reversal occurs when the radial derivative of the bounce action \eqref{eq:Jb_elliptic} changes sign.

\section{\label{sec:sum}Summary}

The analytical formulas presented in this paper provide compact expressions in terms of the Jacobi elliptic functions $({\rm cn}, {\rm sn}, {\rm dn})$ that describe trapped-particle $(\kappa < 1)$ and passing-particle $(\kappa > 1)$ guiding-center orbits in simple axisymmetric tokamak geometry for arbitrary bounce-transit parameter $\kappa$. The parallel coordinates $(s,p_{\|})$ derived in Sec.~\ref{sec:canonical}, which satisfy the canonical condition \eqref{eq:canonical_bounce} for trapped particles and \eqref{eq:canonical_transit} for passing particles, guarantee that the bounce-center transformation \cite{Brizard_bc} in axisymmetric tokamak geometry can be explicitly carried out beyond the deeply-trapped and energetic-passing limits \cite{Wang_Hahm}. 

Applications of the compact Jacobi-elliptic representations of trapped-particle and passing-particle guiding-center orbits include the derivation of bounce-center-averaged fluctuating potentials in bounce-kinetic theory \cite{Brizard_bc,Wang_Hahm} and the derivation of a bounce-center Fokker-Planck collision operator \cite{Brizard_gcFP,Brizard_OAgcFP} beyond the zero-banana-width approximation.


\begin{thebibliography}{5}

\bibitem{Galeev_Sagdeev} A.~A.~Galeev and R.~Z.~Sagdeev, {\it Theory of neoclassical diffusion}, in {\it Reviews of Plasma Physics, vol.~7}, 
M.~A.~Leontovich, ed. (Consultants Bureau, New York, 1979) 257-343.

\bibitem{Hinton_Hazeltine} F.~L.~Hinton and R.~D.~Hazeltine, Rev.~Mod.~Phys.~{\bf 48}, 239 (1976).

\bibitem{anomalous} For a recent review, see E.~J.~Doyle, W.~A.~Houlberg, Y.~Kamada, V.~Mukhovatov, T.~H.~Osborne, A.~Polevoi, G.~Bateman, J.~W.~Connor, J.~G.~Cordey, T.~Fujita, X.~Garbet, T.~S.~Hahm, L.~D.~Horton, A.~E.~Hubbard, F.~Imbeaux, F.~Jenko, J.~E.~Kinsey, Y.~Kishimoto, J.~Li, T.~C.~Luce, 
Y.~Martin, M.~Ossipenko, V.~Parail, A.~Peeters, T.~L.~Rhodes, J.~E.~Rice, C.~M.~Roach, V.~Rozhansky, F.~Ryter, G.~Saibene, R.~Sartori, A.~C.~C.~Sips, 
J.~A.~Snipes, M.~Sugihara, E.~J.~Synakowski, H.~Takenaga, T.~Takizuka, K.~Thomsen, M.~R.~Wade, H.~R.~Wilson, ITPA Transport Physics Topical Group, ITPA Confinement Database and Modeling Topical Group and ITPA Pedestal and Edge Topical Group, Nuc.~Fusion {\bf 47}, S18 (2007). 

\bibitem{GD_90} F.~Y.~Gang and P.~H.~Diamond, Phys.~Fluids B{\bf 2}, 2976 (1990).

\bibitem{Brizard_bc} A.~J.~Brizard, Phys.~Plasmas {\bf 7}, 3238 (2000).

\bibitem{Wang_Hahm} L.~Wang and T.~S.~Hahm, Phys.~Plasmas {\bf 16}, 062309 (2009).

\bibitem{Brizard_gcFP} A.~J.~Brizard, Phys.~Plasmas {\bf 11}, 4429 (2004).

\bibitem{Brizard_OAgcFP} A.~J.~Brizard, J.~Decker, Y.~Peysson, and F.-X.~Duthoit, Phys.~Plasmas {\bf 16}, 102304 (2009).

\bibitem{RBW} R.~B.~White, {\it The Theory of Toroidally Confined Plasmas} (World Scientific, Imperial College Press, 2001).

\bibitem{HMF_elliptic} L.~M.~Milne-Thomson, {\it Elliptic Integrals}, in {\it Handbook of Mathematical Functions}, 
M.~Abramowitz and I.~A.~Stegun, eds. (Dover, New York, 1965) chap.~17.

\bibitem{HMF_Jacobi} L.~M.~Milne-Thomson, {\it Jacobi Elliptic Functions and Theta Functions}, in {\it Handbook of Mathematical Functions}, 
M.~Abramowitz and I.~A.~Stegun, eds.~(Dover, New York, 1965) chap.~16.

\bibitem{Brizard_elliptic} For a recent paper reviewing applications of elliptic functions in classical mechanics, see A.~J.~Brizard, 
Eur.~J.~Phys.~{\bf 30}, 729 (2009).

\bibitem{Duthoit_etal} F.-X.~Duthoit, A.~J.~Brizard, Y.~Peysson, and J.~Decker, Phys.~Plasmas {\bf 17}, 102903 (2010).

\bibitem{footnote} The partial derivative of Jacobi elliptic functions with respect to the modulus $\kappa$ is expressed in terms of the indefinite integral $\int{\rm cn}^{2}(\chi|\kappa)\,d\chi \equiv (1 - \kappa^{-1})\,\chi + \kappa^{-1}{\sf E}(\chi|\kappa)$, where ${\sf E}(\chi|\kappa)$ denotes the incomplete elliptic integral of the second kind \cite{HMF_elliptic}.

\bibitem{Eriksson_Porcelli} L.~G.~Eriksson and F.~Porcelli, Plasma Phys.~Contr.~Fusion {\bf 43}, R145 (2001).

\end{thebibliography}
\end{document}